\documentstyle[12pt]{article}
\def\lapproxeq{\lower .7ex\hbox{$\;\stackrel{\textstyle <}{\sim}\;$}}
\def\gapproxeq{\lower .7ex\hbox{$\;\stackrel{\textstyle >}{\sim}\;$}}

\begin{document}
\titlepage

\begin{flushright}
DTP/94-114 \\ November 1994
\end{flushright}

\vspace*{1cm}

\begin{center}
{\bf CHROMO- AND ELECTRODYNAMICS \\ OF HEAVY UNSTABLE PARTICLES}
\end{center}

\begin{center}
VALERY A.\ KHOZE\footnote{Invited talk at the First Arctic Workshop on Future
Physics and Accelerators, Saariselk\"a, Lapland, Finland, August 21-26,
1994.}\\
Department of Physics, University of Durham, \\ Durham DH1 3LE, U.K.
\end{center}

\vspace*{2cm}

\begin{abstract}
In this talk I attempt to survey some selected physics issues on radiative
interference phenomena in the production of heavy unstable particles.  A
special
emphasis is placed on the reactions $e^+e^- \rightarrow W^+W^- \rightarrow 4$
jets and $e^+e^- \rightarrow t\bar{t} \rightarrow bW^+\bar{b}W^-$.

A transparent recipe is given for quantifying the level of suppression of the
interference effects in the inclusive production processes.  The influence of
the $W$ width on the Coulomb corrections to the threshold $W^+W^-$ production
is
briefly addressed.
\end{abstract}

\newpage

\noindent  {\bf 1.  Introduction}
\vspace*{0.4cm}

There is a common belief that in particle physics \lq\lq tomorrow belongs" to
the detailed studies of heavy unstable objects.  Firstly, we anticipate the
exciting discoveries of new heavy particles (Higgs boson(s), SUSY particles,
$W^{\prime},Z^{\prime}$,...) at increasingly higher energies.  Secondly, for
the
precision tests of the Standard Model one needs the high accuracy determination
of the parameters of the $W$ boson and of the top quark, primarily their
masses.

Let us briefly address the latter point.  This year we have witnessed several
very important developments in precision electroweak tests.  Besides the record
achievements of LEP1 and SLC in measurements of the $Z^0$ parameters$^{[1-3]}$,
one of the most sensational news was the direct evidence reported by the CDF
collaboration for production of the top quark$^{[4]}$.  Meanwhile, all the
dominant loop radiative corrections to the $Z^0$-boson physics are now
available$^{[5,6]}$.  One can consider as an impressive success of the Standard
Model the fact that the top mass $m_t$ predicted from the electroweak data
agrees within the stated errors with the tentative direct CDF observation
\begin{equation}
m_t = 174 \pm 16 \;\; {\rm GeV} .
\end{equation}
There has also been further progress with the determination of the $W$ boson
mass $m_W$ and width $\Gamma_W \,^{[7-9]}$ from the analysis of $W \rightarrow
e\nu$ and $W \rightarrow \mu \nu$ events in $\bar{p}p$ collisions.  The
combined
$W$ mass result gives
\begin{equation}
m_W = 80.23 \pm 0.18 \;\; {\rm GeV} .
\end{equation}
The CDF collaboration has also performed a direct measurement of $\Gamma_W
\,^{[9]}$.  The result is in  good agreement with the indirect measurements and
a Standard Model prediction.   At the same time, so far no strong indication of
physics beyond the Standard Model has been found, and no significant new
constraint on the Higgs mass has been obtained.  To fully exploit the
remarkable
accuracy of exploring the $Z^0$ physics the other precise measurements have to
be performed.  The key role here belongs to the precise determination of $m_W$
and $m_t$.  This could open a new era in the analysis of electroweak data.  One
may hope to pin down the Higgs mass or/and to look for possible manifestations
of physics beyond the Standard Model.

What are the prospects of the experimental studies?  Future improvements in
measurements of $m_W$ are expected from the CDF and D\O detectors at the
Tevatron$^{[7,8]}$.  There are hopes to reduce the $W$ mass uncertainty up to
$\pm$ 100 (or even $\pm$ 50) MeV.  The precise determination of $m_W$ is one of
the main objectives of the upgrade of LEP by a factor 2 in energy, called the
LEP2 program$^{[10,11]}$.  With the foreseen integrated luminosity of 500
pb$^{-1}$ a statistical error of about 60 MeV per experiment is anticipated.

Let  us turn to the top mass determination.  The future studies at the Tevatron
could increase an accuracy of $m_t$ measurements up to $\pm$ 5 GeV.  A unique
precise determination of $m_t$ (with an accuracy of a few hundred MeV) will be
one of the most attractive physics topics at future linear $e^+e^-$
colliders$^{[12-15]}$.

An obvious requirement for success of these precise studies is that the
accuracy
of the theoretical predictions should match or better exceed the experimental
errors.  This requires a detailed understanding of production and decay
mechanisms and, in particular, of the effects arising from the large width,
$\Gamma \sim 0$ (1 GeV).  Recall that in production processes of heavy unstable
particles it is natural to separate the production stage from the decay
processes.  In general these stages are not independent and may be
interconnected by radiative interference effects.  Particle(s) (e.g.\ gluon(s)
and/or photon(s)) could be produced at one stage and absorbed at another; we
speak of virtual interference.  Real interference will occur as well since the
same real particle can be emitted from the different stages of the process.

Many observations rely on a clear understanding of the role of these
interference effects.  Indeed there is a long list of examples where a detailed
knowledge of interferences can be important for the interpretation of
experimental data (see [16-24]).

In this talk I concentrate mainly on the QCD interconnection phenomena that may
occur when two unstable particles ($W$ bosons, top quarks) decay and hadronize
close to each other in space and time.  The word \lq interconnection' is here
introduced to cover those aspects of final-state particle production that are
not dictated by the separate decays of unstable objects, but can only be
understood in terms of the joint action of the two.

Finally, a transparent recipe is presented for quantifying the level of
suppression of the radiative interferences in the inclusive production of heavy
unstable particles.  Some particular attention is paid to the effects of the
$W$
width on the QED Coulomb corrections to $W^+W^-$ production.

\vspace*{0.6cm}

\noindent  {\bf 2.  Colour Reconnection in Hadronic $W^+W^-$ Events}
\vspace*{0.4cm}

The preferable approach to determine the $W$ mass at LEP2 is the method based
on
the direct kinematical reconstruction of $m_W$ in fully hadronic
events$^{[10]}$.  QCD interconnection effects between the $W^+$ and $W^-$ are
of
special practical interest here because they undermine the traditional meaning
of a $W$ mass in the process
\begin{equation}
e^+e^- \rightarrow W^+W^- \rightarrow q_1\bar{q}_2q_3\bar{q}_4 .
\end{equation}
Specifically, it is not even in principle possible to subdivide the final state
into two groups of particles, one of which is produced by the $q_1\bar{q}_2$
system of the $W^+$ decay and the other by the $q_3\bar{q}_4$ system of the
$W^-$ decay: some particles originate from the joint action of the two systems.
Therefore, it is important to understand how large the ambiguities can
be.\footnote{Despite its evident shortcomings (large missing momentum) the
mixed
leptonic-hadronic channel is free from the potential reconnection-induced
ambiguities.}  A systematic analysis of QCD rearrangement phenomena in hadronic
$W^+W^-$ events has been performed in [23] which we follow here.

The perturbative aspects of QCD interference phenomena could be complex but, in
principle, are well controllable.  However, a complete description of these
effects is not possible because of the lack of any deep understanding of the
non-perturbative hadronization process.  Here one has to rely on the model
predictions rather than on exact calculations.  The concept of colour
reconnection (rearrangement/recoupling) is useful to quantify the interference
effects (at least in a first approximation).  In a reconnection two original
colour singlets (such as $q_1\bar{q}_2$ and $q_3\bar{q}_4$) are transmuted into
two new ones (such as $q_1\bar{q}_4$ and $q_3\bar{q}_2$).  Subsequently each
singlet system is assumed to hadronize independently according to the standard
algorithms, which have been so successful in describing e.g.\ $Z^0$ decays.
Depending on whether a reconnection has occurred or not, the hadronic final
state is then going to be somewhat different.

The colour reconnection effects were first discussed by Gustafson, Pettersson
and Zerwas$^{[25]}$, but their results were mainly qualitative and were not
targeted on what might actually be expected at LEP2.  Their picture represents
an example of the so-called instantaneous reconnection scenario, where the
alternative colour singlets are immediately formed and allowed to radiate
perturbative gluons.   Further detailed analysis was performed in Refs.\
[23,26]
with the emphasis on LEP2 studies.

In order to understand which QCD interference effects can occur in hadronic
$W^+W^-$ decays, it is useful to examine the space-time picture of the process.
Consider a typical c.m.\ energy of 170 GeV, a $W$ mass $m_W = 80.2$ GeV and
width $\Gamma_W = 2.07$ GeV.  Then a mean separation of the two decay vertices
in space and in time is of order 0.1 fm.  A gluon with an energy $\omega \gg
\Gamma_W$ therefore has a wavelength much smaller than the separation between
the $W^+$ and $W^-$ decay vertices, and is emitted almost incoherently either
by
the $q_1\bar{q}_2$ system or by the $q_3\bar{q}_4$ one$^{[17]}$.  Only fairly
soft gluons, $\omega \lapproxeq \Gamma_W$, feel the joint action of all four
quark colour charges.  On the other hand, the typical distance scale of
hadronization is about 1 fm, i.e.\ much larger than the decay vertex
separation.
Therefore the hadronization phase may contain significant interference effects.
In the following, we will first discuss perturbative effects and subsequently
non-perturbative ones.

Until recently, perturbative QCD has mainly been applied to systems of primary
partons produced almost simultaneously.  The radiation accompanying such a
system can be represented as a superposition of gauge-invariant terms, in which
each external quark line is uniquely connected to an external antiquark line of
the same colour.  The system is thus decomposed into a set of colourless
$q\bar{q}$ antennae/dipoles$^{[27]}$.  One of the simplest examples is the
celebrated $q\bar{q}g$ system, which (to leading order in $1/N^2_C$, where $N_C
= 3$ is the number of colours) is well approximated by the incoherent sum of
two
separate antennae, $\widehat{qg}$ and $\widehat{g\bar{q}}$.  These dipoles
radiate gluons, which within the perturbative scenario are the principal
sources
of multiple hadroproduction.

Neglecting interferences, the $e^+e^- \rightarrow W^+W^- \rightarrow
q_1\bar{q}_2q_3\bar{q}_4$ final state can be subdivided into two separate
dipoles, $\widehat{q_1\bar{q}_2}$ and $\widehat{q_3\bar{q}_4}$.  Each dipole
may
radiate gluons from a maximum scale $m_W$ downwards.  Within the perturbative
approach, colour transmutations can result only from the interference between
gluons (virtual as well as real) radiated in the $W^+$ and $W^-$ decays.  A
colour reconnection then corresponds to radiation, e.g.\ from the dipoles
$\widehat{q_1\bar{q}_4}$ and $\widehat{q_3\bar{q}_2}$.  The emission of a
single
primary gluon cannot give interference effects, by colour conservation, so
interference terms only enter in second order in $\alpha_s$.

Thus, at least two primary gluons, real or virtual, should be emitted to
generate a colour flow rearrangement, see Figs.\ 1-3.  Note that the diagrams
of
Figs.\ 1a and 1b do not interfere with each other, and that the diagrams of
Fig.\ 2 could interfere with those with single gluon emission, thus inducing a
colour transmutation.  The infrared divergences in the virtual pieces are
cancelled by the corresponding real emissions.  For the case of decay-decay
radiative interference the soft emissions are cancelled in the inclusive cross
section up to at least ${\cal O}(\Gamma_W/m_W$) (see Refs.\ 19,20 and Section
4).

The main qualitative results for the reconnection effects appearing in ${\cal
O}(\alpha^2_s)$ are not much different for various decay-decay interference
samples.  We shall examine below one example corresponding to the diagrams of
Fig.\ 1a.  Let us label the momenta of the final state quarks by $e^+e^-
\rightarrow q_1(p_1)\bar{q}_2(p_2)q_3(p_3)\bar{q}_4(p_4)$ with $Q_1 = p_1 +
p_2$
and $Q_2 = p_3 + p_4$.  In the limit $k_1, k_2 \ll p_i$ after summing over
colours and spins, the interference term may be presented in the form
\begin{equation}
\frac{1}{\sigma_0} d\sigma^{int}_a \simeq \frac{d^3k_1}{\omega_1}
\frac{d^3k_2}{\omega_2} \left( \frac{C_F\alpha_s}{4\pi^2} \right)^2
\frac{1}{N^2_C - 1} \chi_{12} H(k_1) H(k_2) ,
\end{equation}
where $C_F = (N^2_C - 1)/(2N_C) = 4/3$.  We proceed to comment on the
non-trivial factors in this expression.

The radiation pattern $H(k)$ is given by
\begin{equation}
H(k) = \widehat{q_1\bar{q}_4} + \widehat{q_3\bar{q}_2} - \widehat{q_1q_3} -
\widehat{\bar{q}_2\bar{q}_4} ,
\end{equation}
where the radiation antennae are$^{[27]}$
\begin{equation}
\widehat{ij} = \frac{(p_i\cdot p_j)}{(p_i\cdot k)(p_j\cdot k)} .
\end{equation}
The so-called profile function $\chi_{12}$ $^{[16,17]}$ controls decay-decay
radiative interferences
\begin{equation}
\chi_{12} = \left( \frac{m_W\Gamma_W}{\pi} \right) {\cal R} \int dQ^2_1 dQ^2_2
D(Q_1 + k_1) D^*(Q_1 + k_2) D(Q_2 + k_2) D^*(Q_2 + k_1) .
\end{equation}
Here $D(Q)$ is the propagator function
\begin{equation}
D(Q) = \frac{1}{Q^2-m_W^2 + im_W\Gamma_W} ,
\end{equation}
$D^*$ is the complex conjugate of $D$ and ${\cal R}$ represents the real part.
The profile function has the formal property that $\chi_{12} \rightarrow 0$ as
$\Gamma_W \rightarrow 0$ and $\chi_{12} \rightarrow 1$ as $\Gamma_W \rightarrow
\infty$.

The interference is suppressed by $1/(N^2_C-1) = 1/8$ as compared to the total
rate of double primary gluon emissions.  This is a result of the ratio of the
corresponding colour traces,
\begin{equation}
\frac{Tr(T^aT^b) \cdot Tr(T^aT^b)}{Tr(T^aT^a) \cdot Tr(T^bT^b)} =
\frac{(C_FN_C)/2}{(C_FN_C)^2} = \frac{1}{N^2_C-1} .
\end{equation}
Such a suppression takes place for any decay-decay radiative interference
piece,
real as well as virtual, as is clear from Fig.\ 3.

Near threshold and in the limit of massless quarks the interference
contribution
to the radiation pattern is
\begin{equation}
{\cal F}^{int}_a = \frac{2\chi_{12}}{\omega^2_1\omega^2_2} \frac{16 {\rm
cos}\phi_{13} {\rm cos}\tilde{\phi}_{13}}{{\rm sin}\theta_1 {\rm sin}\theta_3
{\rm sin}\tilde{\theta}_1 {\rm sin}\tilde{\theta}_3} ,
\end{equation}
where $\theta_i (\tilde{\theta}_i)$ is the angle between the $q_i$ and the
gluon
$k_1 (k_2)$, and $\phi_{13} (\tilde{\phi}_{13})$ is the relative azimuth
between
$q_1$ and $q_3$ around the direction of the $k_1(k_2)$.  The expression in Eq.\
(10) evidently contains a dependence on the relative orientation of the decay
products of the two $W$'s.  (The interference is maximal when all the partons
lie in the same plane, $\phi_{13} = \tilde{\phi}_{13} = 0$.)  Therefore one
might expect that the decay-decay interferences would induce some
colour-suppressed reconnection effects in the structure of final states in
process (3).

It is the profile function $\chi_{12}$ that cuts down the phase space available
for gluon emissions by the alternative quark pairs (or by any accidental colour
singlets) and thus eliminates the very possibility for the reconnected systems
to develop QCD cascades.  That the $W$ width does control the radiative
interferences can be easily understood by considering the extreme cases.

If the $W$-boson lifetime could be considered as very short, $1/\Gamma_W
\rightarrow 0$, both the $q_1\bar{q}_2$ and $q_3\bar{q}_4$ pairs appear almost
instantaneously, and they radiate coherently, as though produced at the same
vertex.  In the other extreme, $\Gamma_W \rightarrow 0$, the $q_1\bar{q}_2$ and
$q_3\bar{q}_4$ pairs appear at very different times $t_1,t_2$ after the
$W^+W^-$
production,
\begin{equation}
\tau_p \sim \frac{1}{m_W} \ll \Delta t = |t_1 - t_2| \sim \frac{1}{\Gamma_W} .
\end{equation}
The two dipoles therefore radiate gluons and produce hadrons according to the
no-reconnection scenario.

The crucial point is the proper choice of the scale the $W$ width should be
compared with.  That scale is set by the energies of primary emissions, real or
virtual$^{[17,20]}$.  Let us clarify this supposing, for simplicity, that we
are
in the $W^+W^-$ threshold region.  The relative phases of radiation
accompanying
two $W$ decays are then given by the quantity
\begin{equation}
\omega_i \Delta t \sim \frac{\omega_i}{\Gamma_W} .
\end{equation}
When $\omega_i/\Gamma_W \gg 1$ the phases fluctuate wildly and the interference
terms vanish.  This is a direct consequence of the radiophysics of the colour
flows$^{[27]}$ reflecting the wave dynamics of QCD.  The argumentation remains
valid for energies above the $W^+W^-$ threshold as well.

An instructive Gedanken experiment to highlight the filtering role of $\Gamma$
can be obtained by comparing the emission of photons in the eV to MeV range for
the two processes
\begin{equation}
\gamma\gamma \rightarrow W^+W^- \rightarrow \mu^+ \nu_{\mu} \mu^-
\bar{\nu}_{\mu} ,
\end{equation}

\begin{equation}
\gamma\gamma \rightarrow K^+K^- \rightarrow \mu^+ \nu_{\mu} \mu^-
\bar{\nu}_{\mu} ,
\end{equation}
near threshold, in the extreme kinematical configuration where the $\mu^+$ is
collinear with the $\mu^-$.  For the first process, $\omega \ll \Gamma_W$, and
one expects hardly any radiation at all, because of the complete screening of
the two oppositely charged muons.  For the second process, $\omega \gg
\Gamma_K$, the parent particles have long lifetimes and the $\mu^+$ and $\mu^-$
appear at very different times.  The photon wavelength is very small compared
with the size of the $\mu^+\mu^-$ dipole and, therefore, the $\mu^+$ and
$\mu^-$
radiate photons independently, with no interference.

Suppression of the interference in the case of radiation with $\omega_i \gg
\Gamma_W$ can be demonstrated also in a more formal way.

One can perform the integration over $dQ^2_1$ and $dQ^2_2$ in eq.\ (7) by
taking
the residues of the poles in the propagators.  This gives
\begin{equation}
\chi_{12} = \frac{m^2_W\Gamma^2_W(\kappa_1\kappa_2 + m^2_W
\Gamma^2_W)}{(\kappa^2_1 + m^2_W\Gamma^2_W)(\kappa^2_2 + m^2_W\Gamma^2_W)} ,
\end{equation}
with
\begin{equation}
\kappa_{1,2} = Q_{1,2}\cdot (k_1-k_2) .
\end{equation}
For the interference between the diagrams of Fig.\ 1b, the corresponding
profile
function is given by the same formula with $k_2 \rightarrow -k_2$.  Near the
$W^+W^-$ pair threshold Eq.\ (15) is reduced to
\begin{equation}
\chi_{12} = \frac{\Gamma^2_W}{\Gamma^2_W + (\omega_1 - \omega_2)^2} .
\end{equation}

 From Eq.\ (17) it is clear that only primary emissions with $\omega_{1,2}
\lapproxeq \Gamma_W$ can induce significant rearrangement effects: the
radiation
of energetic gluons (real or virtual) with $\omega_{1,2} \gg \Gamma_W$ pushes
the $W$ propagators far off their non-radiative resonant positions, so that the
propagator functions $D(Q_1 + k_1)$ and $D(Q_1 + k_2) (D(Q_2 + k_1)$ and $D(Q_2
+ k_2))$ corresponding to the same $W$ practically do not overlap.  We can
neglect the contribution to the inclusive cross section from kinematical
configurations with $\omega_1,\omega_2 \gg \Gamma_W$, $|\omega_1 - \omega_2|
\lapproxeq \Gamma_W$ since the corresponding phase-space volume is negligibly
small.  The possibility for the reconnected systems to develop QCD cascades is
thus reduced, i.e.\ the dipoles are almost sterile.  Other interferences (real
or virtual) are described by somewhat different expressions, e.g.\ with
$\omega_1 - \omega_2 \rightarrow \omega_1 + \omega_2$, but have the same
general
properties.   Eq.\ (15) clearly shows that $\chi_{12}$ vanishes if any of the
scalar products $Q_i\cdot k_j (i,j = 1,2)$ well exceeds $m_W\Gamma_W$.  Again
accidental kinematics with $\kappa_1,\kappa_2 \ll m_W\Gamma_W$ is suppressed
because of phase space reasons.  Hence all our arguments concerning cutting
down
the QCD cascades induced by the alternative systems remain valid above the
threshold as well.  The smallness of the decay-decay radiative interference for
energetic emission in the production of a heavy unstable particle pair, at the
threshold and far above it, proves to be of a general nature.  For the case of
$e^+e^- \rightarrow bW^+\bar{b}W^-$ this was explicitly demonstrated in [16]
(see also Section 3).

 From the antenna pattern given by Eq.\ (5) one immediately sees that, in
addition to the two dipoles $\widehat{q_1\bar{q}_4}$ and
$\widehat{q_3\bar{q}_2}$, which may be interpreted in terms of reconnected
colour singlets, two other terms, $\widehat{q_1q_3}$ and
$\widehat{\bar{q}_2\bar{q}_4}$ appear.  These terms are intimately connected
with the conservation of colour currents.  Moreover, the $\widehat{q_1q_3}$ and
$\widehat{\bar{q}_2\bar{q}_4}$ antennae come in with a negative sign.  In
general, QCD radiophysics predicts both attractive and repulsive forces between
quarks and antiquarks, see [27-29].  Normally the repulsion effects are quite
small, but in the case of colour-suppressed phenomena they may play an
important
role.

One can elucidate the physical origin of the attraction and repulsion effects
by
examining the photonic interference pattern in
\begin{equation}
\gamma\gamma \rightarrow Z^0Z^0 \rightarrow e^+e^- \mu^+ \mu^- ,
\end{equation}
(see [30]).  In addition to the attractive forces between opposite electrical
charges ($\widehat{e^-\mu^+}$ and $\widehat{e^+\mu^-}$ QED-antennae) there is a
negative-sign contribution $(\widehat{e^-\mu^-}$ and $\widehat{e^+\mu^+}$
QED-antennae) corresponding to the repulsive forces between two same-sign
charges.

It should be emphasized that within the perturbative picture, analogously to
other colour-suppressed interference phenomena$^{[27-29]}$, rearrangement can
be
viewed only on a completely inclusive basis, when all the antennae are
simultaneously active in the particle production.  The very fact that the
reconnection pieces are not positive-definite reflects their wave interference
nature.  Therefore the effects of reconnected almost sterile cascades should
appear on top of a dominant background generated by the ordinary-looking
no-reconnection dipoles $\widehat{q_1\bar{q}_2}$ and $\widehat{q_3\bar{q}_4}$.

Summing up the above discussion, it can be concluded that perturbative colour
reconnection phenomena are suppressed, firstly because of the overall factor
$\alpha^2_s/(N^2_C - 1)$, and secondly because the rearranged dipoles can only
radiate gluons with energies $\omega \lapproxeq \Gamma_W$.  Only a few
low-energy particles should therefore be affected, \linebreak  $\Delta
N^{recon}/N^{no-recon} \lapproxeq {\cal O}(10^{-2})$.

Having demonstrated that perturbative rearrangement is very small we now turn
to
the possibility of reconnection occuring as a part of the non-perturbative
fragmentation phase.  Since hadronization is not well understood, this requires
model building.  In [23] the Standard Lund string fragmentation model$^{[31]}$
was used as a starting point\footnote{This choice, by no means, was dictated by
a tendency to ignore the other successful hadronization schemes but rather by
personal experience.  Recall that all the so-called WIG'ged (\underline{W}ith
\underline{I}nterfering \underline{G}luons) Monte-Carlo algorithms (HERWIG,
JETSET, ARIADNE) describe very successfully a wealth of the existing
experimental data.  An examination within the framework of other fragmentation
models looks rather useful.} but it was considerably extended.

Recall that the string description is entirely probabilistic, i.e.\ any
negative-sign interference effects are absent.  This means that the original
colour singlets $q_1\bar{q}_2$ and $q_3\bar{q}_4$ may transmute to new singlets
$q_1\bar{q}_4$ and $q_3\bar{q}_2$, but that any effects, e.g., of the $q_1q_3$
and $\bar{q}_2\bar{q}_4$ dipoles (cf.\ Eq.\ (5)) are absent.  In this respect,
the non-perturbative discussion is more limited in outlook than the
perturbative
one above.  This does not necessarily mean that there is a physics conflict
between the two pictures: one should remember that the perturbative approach
describes short-distance phenomena, where partons may be considered free to
first approximation, while the Lund string picture is a model for the
long-distance behaviour of QCD, where confinement effects should lead to a
subdivision of the full system into colour singlet subsystems (ultimately
hadrons) with screened interactions between these subsystems.\footnote{However,
in my view, the lack of understanding of how to handle the negative sign
antenna
pieces casts some shadow on the reliability of the probabilistic descriptions.
I prefer to consider the latter as a reasonable qualitative guide allowing one
to estimate the size of the interconnection effects rather than the complete
predictions.}

The imagined time sequence is the following (for details see [30]).  The $W^+$
and $W^-$ fly apart from their common production vertex and decay at some
distance.  Around each of these decay vertices, a perturbative parton shower
evolves from an original $q\bar{q}$ pair.  The typical distance that a virtual
parton (of mass $m \sim 10$ GeV, say, so that it can produce separate jets in
the hadronic final state) travels before branching is comparable with the
average $W^+W^-$ separation, but shorter than the fragmentation time.  Each $W$
can therefore effectively be viewed as instantaneously decaying into a string
spanned between the partons, from a quark end via a number of intermediate
gluons to the antiquark end.  The strings expand, both transversely and
longitudinally.  They eventually fragment into hadrons and disappear.  Before
that time, however, the string from the $W^+$ and the one from the $W^-$ may
overlap.  If so, there is some probability for a colour reconnection to occur
in
the overlap region.  The fragmentation process is then modified.

The Lund string model does not constrain the nature of the string fully.  At
one
extreme, the string may be viewed as an elongated bag, i.e.\ as a flux tube
without any pronounced internal structure.  At the other extreme, the string
contains a very thin core, a vortex line, which carries all the topological
information, while the energy is distributed over a larger surrounding region.
The latter alternative is the chromoelectric analogue to the magnetic flux
lines
in a type II superconductor, whereas the former one is more akin to the
structure of a type I superconductor.  One can use them as starting points for
two contrasting approaches, with nomenclature inspired by the superconductor
analogy.

In scenario I, the reconnection probability ${\cal P}_{recon}$ is proportional
to the space-time volume over which the $W^+$ and $W^-$ strings overlap, with
strings assumed to have transverse dimensions of hadronic size.  In scenario II
it is assumed that reconnections can only take place when the core regions of
two string pieces cross each other.  This means that the transverse extent of
strings can be neglected, which leads to considerable simplification.

Both scenarios were implemented in a detailed simulation of the full process of
$W^{\pm}$ production and decay, parton shower evolution and
hadronization$^{[32]}$.  It is therefore possible to assess any experimental
consequences for an ideal detector.

The reconnection probability is predicted in scenario II without any adjustable
parameters, although with the possibility to vary the baseline model in a few
respects.  Scenario I contains a completely free strength parameter $k_1$.  It
was chosen to give an average  ${\cal P}_{recon} \approx 0.35$ at 170 GeV, as
is
predicted  in scenario II.

An instructive issue is related to the energy dependence of the reconnection
phenomena.  At first glance, one might expect that these \lq\lq undesirable"
effects \lq\lq go away" with an increase of the threshold energy.  However, as
was demonstrated in [23] the resulting c.m.\ energy dependence of ${\cal
P}_{recon}$ in the whole LEP2 region is very slow: between 150 and 200 GeV the
variation is less than a factor of 2.  Here it is useful to remember that the
$W^{\pm}$ are never produced at rest with respect to each other: the naive
Breit-Wigner mass distributions are distorted by phase-space effects, which
favour lower $W$ masses.\footnote{Similar effects appear in the momentum
distribution of the top quarks in the threshold region (see [33]).  Here they
are additionally modified by the final state QCD interactions.}  For 150-200
GeV
the average momentum of each $W$ is therefore in the range 22-60 GeV, rather
than in the range 0-60 GeV, see Fig.\ 4.  It is largely this momentum that
indicates how fast the two $W$ system are flying apart, and therefore how much
they overlap in the middle of the event.  Also the energy variation in the
perturbative description is very small.  If we want to call colour reconnection
a threshold effect, we have to acknowledge that the threshold region is very
extended.

Let us turn now to the $W$ mass determination at LEP2.  Experimentally, $m_W$
depends in a non-trivial fashion on all particle momenta of an event.  Errors
in
the $W$ mass determination come from a number of sources$^{[10,11,23]}$ which
we
do not intend to address here.  Therefore, we only study the extent to which
the
average reconstructed $W$ mass is shifted when reconnection effects are added,
but everything else is kept the same.  Even so, results do depend on the
reconstruction algorithm used.  In [23] we have tried a few different ones,
which, however, all are based on the same philosophy: a jet finder is used to
define at least four jets, events with two very nearby jets or with more than
four jets are rejected, the remaining jets are paired to define the two $W$'s,
and the average $W$ mass of the event is calculated.  Events where this number
agrees to better than 10 GeV with the input average mass are used to calculate
the systematic mass shift.

In scenario I this shift is consistent with being zero, within the 10 MeV
uncertainty in our results from limited Monte Carlo statistics (160,000 events
per scenario).  Scenario II gives a negative mass shift, of about -30 MeV; this
also holds for several variations of the basic scheme.  A simpler model, where
reconnections are always assumed to occur at the centre of the event, instead
gives a positive mass shift: about +30 MeV if results are rescaled to ${\cal
P}_{recon} \approx 0.35$.  We are, therefore, forced to conclude that not even
the sign of the effect can be taken for granted, but that a real uncertainty of
$\pm$30 MeV does exist from our ignorance of non-perturbative reconnection
effects.

To estimate the size of perturbative rearrangement, one can use a scenario
where
the original $q_1\bar{q}_2$ and $q_3\bar{q}_4$ dipoles are instantaneously
reconnected to $q_1\bar{q}_4$ and $q_3\bar{q}_2$ ones, and these are allowed to
radiate gluons with an upper cut-off given by the respective dipole invariant
mass$^{[25]}$.  This gives a mass shift by about +500 MeV.  We have above
argued
that real effects would be suppressed by at least a factor of $10^{-2}$
compared
to this, and thus assign a 5 MeV error from this source.  Finally, the
possibility of an interplay between the perturbative and non-perturbative
phases
must be kept in mind.  We believe it will not be much larger than the
perturbative contribution, and thus assign a further 5 MeV from this source.
The numbers are added linearly to get an estimated total uncertainty of 40 MeV.

In view of the aimed-for precision, 40 MeV is non-negligible.  However,
remember
that as a fraction of the $W$ mass itself it is a half a per mille error.
Reconnection effects are therefore smaller in the $W$ mass than in many other
observables, such as the charged multiplicity.  This is good news.  Otherwise,
LEP2 would not have significant advantages in the measurements of $m_W$ over
hadronic colliders where the accuracy is steadily improving.

In total, our conservative estimate of the systematic uncertainty on the $W$
mass would thus be a number roughly like 40 MeV from the reconnection
phenomenon
alone.  It may well be the largest individual source of systematic error for
$W$
mass determinations in doubly hadronic $W^+W^-$ decays.

We cannot today predict what will exactly come out of the forthcoming studies
and the results of [23] might form only a starting point for future
activity.\footnote{A new analysis based on these results is now performed by
the
DELPHI collaboration$^{[11]}$.}  First of all, it is important to study how
sensitive experimental mass reconstruction algorithms are, and not just rely on
the numbers in [23].  We believe that the uncertainty can be reduced by a
suitable tuning of the algorithms, e.g.\ with respect to the importance given
to
low-momentum particles (for which the detection efficiency may be limited) and
with respect to the statistical treatment of the wings of the $W$ mass
distribution.

There is another challenging reason to study the phenomenon of colour
rearrangement in hadronic $W^+W^-$ events.  As was first emphasised in [25], it
could provide a new laboratory for probing the non-perturbative QCD dynamics.
The very fact that different models for the colour string give different
predictions means that it might be possible to learn about the structure of the
QCD vacuum.  For example, one may hope to distinguish scenarios I and II by
exploiting the difference in the sensitivity of the reconnection to the event
topology$^{[23,26]}$.

Unfortunately to make any progress at all, we have had to rely on models and
approximations that are far from perfect.  There is a true limit to our current
physics understanding.  One unresolved problem concerns an evident breakdown of
the exclusive probabilistic interpretation of the colour-suppressed
interference
effects (see [23] and the discussion above).  Another open question addresses
the role of an interplay between the perturbative and non-perturbative
phases$^{[30]}$.

Let us now come to the issue of experimental observability of the reconnection.
An analysis in [23] concerned mainly the standard global event measures where
effects seem to be very small.  The change in the average charged multiplicity
is expected at the level of a percent of less, and similar statements hold for
rapidity distributions, thrust distributions, and so on.  This is below the
experimental precision one may expect at LEP2, and so may well go unobserved.

There are some other potentially promising approaches, e.g.\ comparison of the
event properties in fully hadronic and mixed leptonic-hadronic decays.  An
interesting vista on the reconnection problem is connected with Bose-Einstein
effects$^{[30]}$.

A high-statistics run above the $Z^0Z^0$ threshold would allow an unambiguous
determination of any systematic mass shift, given that the $Z^0$ mass is
already
known from LEP1 with a record accuracy ($\pm$4 MeV$^1$).  If the various
potential sources of systematic error could be disentangled, it could also
imply
a direct observation of reconnection effects.  More generally, $Z^0$ events
from
LEP1 can be used to predict a number of properties for $Z^0Z^0$ events, such as
the charged multiplicity distribution.  Any sign of deviations would provide
important information on the reconnection issue.

A new proposal attempting to disentangle the recoupling phenomenon is discussed
in [26].\footnote{The starting points in this scenario are different from those
in [23].  In particular, it does not use any space-time overlaps and the
time-retardation effects.}  It is advocated there that the dynamical effects
could enhance the reconnection probability for configurations which correspond
to so-called short strings producing few hadrons.  It is predicted that with
10\% probability for recoupling the reconnected events can be experimentally
identified.

In my view, in order to pin down the reconnection in real-life experiments some
further efforts are needed.  Besides all the theoretical uncertainties, the
background issue should be addressed more carefully.  For example, the short
string states could be generated also by the conventional $e^+e^- \rightarrow$
4
parton events (e.g.\ the so-called rapidity gap events$^{[34]}$).  This could
provide a natural lower limit for a recoupling-type signal.

Finally, one may anticipate that the reconnection-type effects might appear
also
in $Z^0$ decay events.  They would be induced, e.g.\ by some subsets of gluons
in the colour singlet states.  It could be quite instructive to test the
predictions of different models for recoupling in hadronic $Z^0$ events, where
one can benefit from the huge statistics of LEP1.

\vspace*{0.6cm}

\noindent  {\bf 3.  Correlations of Particle Flow in Top Events}
\vspace*{0.4cm}

One of the main objectives of a future linear $e^+e^-$ collider will be to
determine the top mass $m_t$ with high accuracy.  Besides the traditional
measurements of the $t\bar{t}$ excitation curve, several other approaches are
discussed$^{[21,22,35,36]}$.  One method is to reconstruct the top invariant
mass event by event, another is to measure the top momentum
distribution$^{[33,35]}$.  In either case, the QCD interconnection effects
could
introduce the potentiality for a systematic bias in the top mass determination.

It is not my intention to go here through all the details of the problem.  As a
specific topical example, we consider the production and decay of a $t\bar{t}$
pair in the process
\begin{equation}
e^+e^- \rightarrow t\bar{t} \rightarrow bW^+\bar{b}W^-
\end{equation}
and concentrate on the possible manifestations of the interconnection effects
in
the distribution of the particle flow in the final state (for details see
[24]).
For simplicity we assume that the $W$'s decay leptonically, so the colour flow
is generated only by the $t$ and $b$ quarks.  Further, we restrict ourselves to
the region a few GeV above the $t\bar{t}$ threshold to exemplify the size of
effects.  The radiation pattern is especially simple in this region.

Recall that the dominance of the $t \rightarrow bW^+$ decay mode leads to a
large top width $\Gamma_t$, which is about 1.3 GeV for a canonical mass $m_t
\simeq 170$ GeV.  This width is larger than the typical hadronic scale $\mu
\sim
1$ fm$^{-1}$, and the top decays before it have time to hadronize$^{[37,38]}$.
It is precisely the large width that makes top physics so unique.  Firstly, the
top decay width $\Gamma_t$ provides an infrared cut-off for the strong forces
between the $t$ and the $\bar{t} \, ^{[39-41]}$.  The width $\Gamma_i$ acts as
a
physical \lq smearing'$^{[42]}$, and the top production becomes a quantitative
prediction of perturbative QCD, largely independent of non-perturbative
phenomenological algorithms.  Secondly, $\Gamma_t$ controls the QCD
interferences between radiation occurring at different stages of the $t\bar{t}$
production processes$^{[16-18]}$.  These interferences affect the structure of
the colour flows in the $t\bar{t}$ events and may provide a potentially serious
source of uncertainties in the reconstruction of the final state.

The interplay of several particle production sources is reminiscent of the
colour rearrangement effects we have studied for process (3), but there are
important differences.  From the onset, $W^+W^-$ events consist of two separate
colour singlets, $q_1\bar{q}_2$ and $q_3\bar{q}_4$, so that there is no logical
imperative of an interplay between the two.  Something extra has to happen to
induce a colour rearrangement to $q_1\bar{q}_4$ and $q_3\bar{q}_2$ singlets,
such as a perturbative exchange of gluons or a non-perturbative string overlap.
This introduces a sizeable dependence on the space-time picture, i.e.\ on how
far separated the $W^+$ and $W^-$ decay vertices are.  The process (19) only
involves one colour singlet.  Therefore an interplay is here inevitable, while
a
colour rearrangement of the above kind is impossible.  Recall also that,
contrary to the $WW$ case, there are no purely leptonic channels which could
provide a reconnection-free environment.  Analogously to the $W^+W^-$ case we
expect that the perturbative restructuring is suppressed because of the
space-time separation between the decays of the $t$ and $\bar{t}$ quarks.
However, a priori there is no obvious reason why interconnection effects have
to
be small in the fragmentation process.  Moreover, the $b$ and $\bar{b}$ coming
from the top decays carry compensating colour charges and therefore have to \lq
cross-talk' in order to produce a final state made up of colourless
hadrons$^{[24,43]}$.

Let us start from the perturbative picture.  In the process (19) the standard
parton showering can be generated by the systems of quarks appearing within a
short time scale, namely the $\widehat{t\bar{t}}, \, \widehat{tb}$ and
$\widehat{\bar{t}\bar{b}}$ antennae/dipoles.  In the absence of interferences
these antennae do not interact and the particle flows are not rearranged.

As was discussed in [16,17], the energy range of primary gluons, real or
virtual, generated by the alternative quark systems of the type
$\widehat{t\bar{b}}, \, \widehat{\bar{t}b}$ and $\widehat{b\bar{b}}$ is
strongly
restricted.  Not so far from the $t\bar{t}$ threshold one expects $\omega
\lapproxeq \omega^{int}_{max} \sim \Gamma_t$.  Therefore the would-be parton
showers initiated by such systems are almost sterile, and can hardly lead to a
sizeable restructuring of the final state.  In other words, the width of an
unstable particle acts as a kind of filter, which retains the bulk of the
radiation (with $\omega > \Gamma_t$) practically unaffected by the relative
orientation of the daughter colour charges.

The general analysis of soft radiation in process (19) in terms of QCD antennae
was presented in [16].  Here we focus on the emission close to the $t\bar{t}$
threshold.

The primary-gluon radiation pattern can be presented as:
\begin{equation}
dN_g \equiv \frac{d\sigma_g}{\sigma_0} = \frac{d\omega}{\omega}
\frac{d\Omega}{4\pi} \frac{C_F\alpha_s}{\pi} {\cal I} ,
\end{equation}
where $\Omega$ denotes the gluon solid angle; ${\cal I}$ is obtained by
integrating the absolute square of the overall effective colour current over
the
virtualities of the $t$ and $\bar{t}$.

Near threshold the contribution from the $\widehat{t\bar{t}}$ antenna may be
neglected, and the $\widehat{tb}$ and $\widehat{\bar{t}\bar{b}}$ antennae are
completely dominated by the emission off the $b$ quarks.  The distribution
${\cal I}$ may then be presented in the form
\begin{equation}
{\cal I} = {\cal I}_{indep} + {\cal I}_{dec-dec} .
\end{equation}
Here ${\cal I}_{indep}$ describes the case when the $b$ quarks radiate
independently and ${\cal I}_{dec-dec}$ corresponds to the interference between
radiation accompanying the decay of the top and of the antitop
\begin{equation}
{\cal I}_{dec-dec} = 2 \chi (\omega) \frac{{\rm cos}\theta_1 {\rm cos}\theta_2
-
{\rm cos}\theta_{12}}{(1 - v_b{\rm cos}\theta_1)(1 - v_{\bar{b}} {\rm
cos}\theta_2)} .
\end{equation}
Here $\theta_1(\theta_2)$ is the angle between the $b(\bar{b})$ and the gluon,
$\theta_{12}$ is the angle between the $b$ and $\bar{b}$ and $\chi(\omega)$ is
the profile function [16], which controls the radiative interferences between
the different stages of process (19).

Near threshold
\begin{equation}
\chi(\omega) = \frac{\Gamma^2_t}{\Gamma^2_t + \omega^2}
\end{equation}
(cf.\ Eq.\ (17)).

Analogously to the $W^+W^-$ case,  the profile function $\chi(\omega)$ cuts
down
the phase space available for emissions by the alternative quark systems and,
thus, suppresses the possibility for such systems to develop QCD cascades.  As
$\Gamma_t \rightarrow \infty$, the top lifetime becomes very short, the $b$ and
$\bar{b}$ appear almost instantaneously, and they radiate coherently, as though
produced directly.  In particular, gluons from the $b$ and $\bar{b}$ interfere
maximally, i.e.\ $\chi(\omega) = 1$.  At the other extreme, for $\Gamma_t
\rightarrow 0$, the top has a long lifetime and the $b$ and $\bar{b}$ appear in
the course of the decays of top-flavoured hadrons at widely separated points in
space and time.  They therefore radiate independently, $\chi(\omega) = 0$.
Thus
a finite top width suppresses the interference compared to the naive
expectation of fully coherent emission.  The same phenomena appear for the
interference contributions corresponding to virtual diagrams.  The infrared
divergences induced by the unobserved gluons are cancelled when both real and
virtual emissions are taken into account.

The bulk of the radiation caused by primary gluons with $\omega > \Gamma_t$ is
governed by the $\widehat{tb}$ and $\widehat{\bar{t}\bar{b}}$ antennae.  It is
thus practically unaffected by the relative orientation of the $b$ and
$\bar{b}$
jets.  In particular, the $\widehat{b\bar{b}}$ antenna is almost inactive.  The
properties of individual $b$ jets are understood well enough, thanks to our
experience with $Z^0 \rightarrow b\bar{b}$ at LEP1.

Because of the suppression of energetic emission associated with the
interferences, the restructuring could affect only a few particles.

Interconnection phenomena could affect the final state of $t\bar{t}$ events in
many respects, but multiplicity distributions are especially transparent to
interpret.  As a specific example, we examined in Ref.\ [24] the total
multiplicity of double leptonic top decays as a function of the relative angle
between the $b$ and $\bar{b}$ jets.  Let us make some comments concerning the
basic ideas of these studies:
\begin{enumerate}
\item As usual, one needs to model the fragmentation stage and study quantities
accessible at the hadron level.  This offers one advantage: the fragmentation
has many similarities with the $\omega \rightarrow 0$ limit of the perturbative
picture, and thus tends to enhance non-trivial angular dependences.
\item  A complication of attempting a full description is that it is no longer
enough to give the rate of primary-gluon emission, as in Eq.\ (20): one must
also allow for secondary branchings and specify the colour topology and
fragmentation properties of radiated partons.  It is then useful to benefit
from
the standard parton shower plus fragmentation picture for $e^+e^- \rightarrow
\gamma^*/Z^0 \rightarrow q\bar{q}$, where these aspects are understood, at
least
in the sense that much of our ignorance has been pushed into experimentally
fixed parameters.
\item  The relation between $\gamma^*/Z^0 \rightarrow q\bar{q}$ and $t\bar{t}
\rightarrow bW^+\bar{b}W^-$ is most easily formulated in the antenna/dipole
language.  The independent emission term corresponds to the sum of two dipoles,
${\cal I}_{indep} \propto \widehat{tb} + \widehat{\bar{t}\bar{b}}$, while the
decay-decay interference one corresponds to ${\cal I}_{dec-dec} \propto
\chi(\omega)(\widehat{b\bar{b}} - \widehat{tb} - \widehat{\bar{t}\bar{b}}$).
In
total, therefore,
\begin{equation}
{\cal I} \propto (1 - \chi(\omega)) \widehat{tb} + (1 - \chi(\omega))
\widehat{\bar{t}\bar{b}} + \chi(\omega) \widehat{b\bar{b}} .
\end{equation}
Each term here is positive definite and can be translated into a recipe for
parton shower evolution: starting from the respective normal branching picture,
each potential primary branching $q \rightarrow qg$ or $\bar{q} \rightarrow
\bar{q}g$ is assigned an additional weight factor $1 - \chi(\omega)$ or
$\chi(\omega)$.  This factor enters the probability that a trial branching will
be retained.  For the rest, the same evolution scheme can be used as for
$\gamma^*/Z^0 \rightarrow q\bar{q}$, including the choice of evolution
variable,
$\alpha_s$ value, and so on.  To first approximation, this means that the
$\widehat{tb}$ and $\widehat{\bar{t}\bar{b}}$ dipoles radiate normally for
$\omega \gapproxeq \Gamma_t$ and have soft radiation cut off, with the opposite
for the $\widehat{b\bar{b}}$ dipole.
\item The top quarks are assumed to decay isotropically in their respective
rest
frame, i.e.\ we do not attempt to include spin correlations between $t$ and
$\bar{t}$.  Close to threshold this gives an essentially flat distribution in
cos$\theta_{parton}$, defined as the angle between the \lq original' $b$ and
$\bar{b}$ directions before QCD radiation effects are considered.  Breit-Wigner
distributions are included for the top and $W$ masses.
\end{enumerate}

On the phenomenological side, the main conclusions of the analysis in [24] are:
\begin{itemize}
\item The interconnection should be readily visible in the variation of the
average multiplicity as a function of the relative angle between the $b$ and
$\bar{b}$ (see Fig.\ 5).
\item  A more detailed test is obtained by splitting the particle content in
momentum bins.  The high-momentum particles are mainly associated with the
$\widehat{tb}$ and $\widehat{\bar{t}\bar{b}}$ dipoles and therefore follow the
$b$ and $\bar{b}$ directions, while the low-momentum ones are sensitive to the
assumed influence of the $\widehat{b\bar{b}}$ dipole.
\item  A correct description of the event shapes in top decay, combined with
sensible reconstruction algorithms, should give errors on the top mass that are
significantly less than 100 MeV.
\end{itemize}

The possibility of interference reconnection effects in $t\bar{t}$ production
is
surely not restricted to the phenomena discussed here.  They could affect
various other processes/characteristics.

One topical example concerns the top quark momentum reconstruction.  As was
emphasised in Ref.\ [33], the momentum measurement combined with the threshold
scan could significantly improve the overall precision in determination of
$m_t$
and  $\Gamma_t$.  As a supplementary bonus, the top momentum proves to be less
sensitive to the beam effects.\footnote{Experimental aspects are summarized,
e.g., in Ref.\ [35].  It is anticipated that $m_t$ and $\frac{\Delta
\Gamma_t}{\Gamma_t}$ will be measured with accuracies of $\approx 350$ MeV and
0.07 respectively.  (I am grateful to K.\ Fujii for discussion of these
results.)}

In order to reconstruct the top momentum we need at least one of the secondary
$W$'s decaying hadronically.  So the final state configurations are either a
lepton plus four quark jets or six quark jets.

QCD interconnection may efface the separate identities of the top and antitop
systems and, thus, could produce a potential source of the systematic error in
the top momentum determination.\footnote{The reconstruction of the $W$-momentum
could be affected as well.  The QCD interferences can have some impact also on
the top momentum distribution itself.  The estimates show that there is,
indeed,
a small effect$^{[20,21]}$.}  The interference pattern here is more complicated
than in the case of  double leptonic decays because of an additional
cross-talking between the hadronically decaying $W$ and the $b\bar{b}$
products.

It is clear that the problem of these interconnection-related uncertainties
deserves further systematic studies.  It may well happen that these
uncertainties are non-negligible in view of the aimed-for high precision of
measurements.

Finally, let us make some comments concerning the definition of $m_t$.  It is
obvious that the high aimed-for precision of the top mass determination at a
future linear collider ($\Delta m_t \approx \Lambda_{QCD}$) requires clear
theoretical understanding of this issue.  After all, a quark is a colour object
surrounded by the (long-distance) colour fields and its physical mass cannot be
unambiguously defined in the full theory once non-perturbative QCD effects are
taken into account.\footnote{For an intensive recent discussion of this old
(but
still hot) topic see, e.g., Refs.\ [44-46].}

To avoid confusion, I would like to stress that it is the so-called physical or
pole mass, $m^p_t$, that is observable experimentally, see also the discussion
in [47].  The pole mass is related to the location of the singularity of the
renormalized quark propagator and appears naturally in the perturbative
calculations.  Unlike many other notions, $m^p_t$ can be defined in a gauge
invariant way.

The pole mass is related to the running $\overline{MS}$ mass evaluated at the
$m^p_t$ scale, $\hat{m}_t \equiv \hat{m}_t(m^p_t)$ as
\begin{equation}
m^p_t = \hat{m}_t \left[ 1 + C_F \frac{\alpha_s(m^p_t)}{\pi} + K_t \left(
\frac{\alpha_s(m^p_t)}{\pi} \right)^2 + ... \right]
\end{equation}
with the two-loop coefficient $K_t$ calculated in Ref.\ [48].  Numerically,
$K_t
\simeq 11$.  Let us point attention to the fact that the difference between
these two masses is quite large (about 10 GeV).

Distinguishing between various masses is an important issue in the higher-order
calculations of various characteristics of the top quark physics.  Each time
one
needs to understand clearly what mass definition is the most appropriate to the
phenomenon in question.  The principal ambiguity in the definition of the pole
mass of a quark$^{[44,45]}$ is deeply rooted in the divergence of perturbation
theory at high orders (the so-called infrared-renormalon problem$^{[49]}$).
Because of the asymptotic explosion of the perturbation series in Eq.\ (25),
truncating it at some optimal order leads to an intrinsic uncertainty of order
$\Lambda_{QCD}$ assuming $\hat{m}_t$ as given$^{[44,45]}$.  This imposes a
limit
($O(\Lambda_{QCD})$) on the precision of the definition of the quark pole
mass.\footnote{In terms of the non-relativistic description of the
quark-antiquark system this long-distance uncertainty is related to the
ambiguity in the additive constant term in the QCD potential, see Refs.\
[36,50].}  On the other hand, the running mass $m_t(\mu)$ does not suffer from
the infrared-renormalon disease and, in principle, can be found to any accuracy
at a high scale $\mu \gg \Lambda_{QCD}$.  Because the top quarks contributing
to
the electroweak loops have high virtualities, the QCD effects are characterized
by momenta of order of $m_t$ and the mass parameter most relevant to the
electroweak corrections (like $\Delta\rho$) is $m_t(m_t) \,^{[46,51]}$.  The
various theoretical expectations for the top quark mass (e.g.\ in SUSY GUTs,
supergravity or superstring models$^{[52]}$) are also given in terms of this
running mass.

An experienced reader may well wonder what is the possible impact of the very
rapid top decay on the issue of its mass.  Recall that the intrinsic
uncertainty
in the pole mass of a heavy quark results from the infrared effects associated
with the gluon momenta $k \sim \Lambda_{QCD}$, and measuring of this mass
requires a very long time, $t_{meas} \sim \frac{1}{\Lambda_{QCD}} \, ^{[44]}$.
This time is of the same order as the infrared gluon formation time, $t_f$.
The
top quark decays well before these gluons can be formed and its physics should
not, in principle, suffer from the long-distance problems.  The infrared
renormalon disease is inherited by the daughter $b$-quarks.

An instructive example comes from examining the soft gluon emission in the
process (19), see [16].  At $k < \Gamma_t$ the radiation decouples from the top
quarks and the infrared effects become associated with the $b\bar{b}$ system
only.

The presence of the width provides further options for the definition of the
on-shell mass (see [47] and references therein).  It appears to be quite
challenging to find a way to reduce the long-distance uncertainties in the
physical mass of the top quark.  We are working now in this direction.

\vspace*{0.6cm}

\noindent {\bf 4.  Radiative Interferences in the Inclusive Production of Heavy
Unstable Particles}

\vspace*{0.4cm}

As we have already discussed, in production processes of heavy unstable
particles it is natural to separate the production stage from the decay stages.
The various production-decay and decay-decay radiative interferences for a
given
process might, in principle, be expected to produce a complicated effect.
Fortunately, as was shown explicitly in [19,20] such interference corrections
are cancelled in the inclusive cross-sections up to terms of relative order
$\alpha_{int} \frac{\Gamma}{M}$ (where $\alpha_{int} = \alpha$ or $\alpha_s$ as
appropriate) or better.  The only exception is the contribution arising from
the
universal Coulomb interaction between slowly moving produced objects.

In [20] an explicit study was performed with the help of soft-insertion
techniques and, in each case, we identified the particular degree of
inclusiveness that is required for the interference effects to be suppressed.
Of course, the real and virtual interference contributions, taken separately,
are not suppressed but are infrared divergent.  Clearly the infrared divergent
parts have to cancel for physically meaning values.  This is not the issue.
Rather the crucial question is to what level does this cancellation occur?

Here, following [19], we present a general theorem which states that the
effects
of radiative interferences are each suppressed by $O(\Gamma/M)$  in the totally
inclusive production.  This proof does not rely on specific assumptions, like
soft-insertion factorization, and the resulting approach is applicable to any
order in $\alpha_{int}$.

It may be useful to express the resulting recipe in a symbolic form.  To be
specific let us consider the production of $N$ heavy unstable particles
$A_1,...A_N$ with masses $M_i$ and widths $\Gamma_i$.  For reference purposes
we
first consider the production in the absence of the decays.  Then the inclusive
cross section may be written in the form
\begin{equation}
\sigma_{stable}(A_1,...A_N) = \sigma_0(M^2_i)(1 + \delta(R,C)) ,
\end{equation}
where $\sigma_0(M^2_i)$ is the production cross section in the Born
approximation and $\delta (R,C)$ represents the radiative corrections.  Here we
have separated Coulomb corrections $C$ from the remaining radiative corrections
$R$.  The Coulomb effects $C$, which are associated with large space-time
intervals, are only important$^{[53]}$ if two charged (or coloured) particles
are slowly moving in their c.m.\ frame.  Note that the separation of Coulomb
effects can only be done uniquely near threshold, but this is the very region
where the instability effects are most important.  As was demonstrated in
[39-41,54] these effects are especially important for process (19) where they
drastically modify the threshold cross section.  We shall briefly address below
the issue of the QED Coulomb corrections to $e^+e^- \rightarrow W^+W^-$, see
Refs.\ [55,56].

The question is \lq\lq how is (26) modified when we allow the particles $A_i$
to
decay?"  We will show
\begin{equation}
\sigma_{unstable}(A_1,...A_N) = \int \prod_i (ds_i\rho (s_i)) \sigma_0(s_i)(1 +
\delta (R,\bar{C})) + \sum_n O \left( \alpha^n_{int} \frac{\Gamma_i}{M_i}
\right)
\end{equation}
with
\begin{equation}
\rho(s_i) = \frac{\sqrt{s_i}\Gamma_i(s_i)}{\pi [(s_i-M^2_i)^2 +
s_i\Gamma^2_i(s_i)]}
\end{equation}
where $\Gamma_i(s_i)$ is \lq\lq running" physical width which incorporates the
radiative effects associated solely with the decay of $A_i$.

In other words, the theorem for the production of heavy {\it unstable}
particles
says that, apart from the two modifications explicitly shown in the formula,
the
introduction of the widths gives rise to no new corrections up to order
$\alpha^n_{int}\Gamma_i/M_i$, where the $n = 0$ term corresponds to the
standard
(non-radiative) non-resonant backgrounds and the $n \geq 1$ terms to
interference induced by $n$ radiated quanta.  The two modifications in going
from (26) to (27) are, first, the natural kinematic effect leading to the
integrations over $\rho(s_i)$ and, second, the modification (symbolically $C
\rightarrow \bar{C}$) of the Coulombic interaction between particle pairs which
are non-relativistic in their c.m.\ frame.  In particular, the theorem says the
remaining radiative corrections are unchanged (see also Ref.\ [57]).

The proof$^{[19]}$ relies on two facts.  First, the suppression of
interferences
between the various production and decay stages arising from energetic
photons/gluons with $|k^0| \gg \Gamma$, and, second, the absence of infrared
divergences in the total or inclusive cross-section.  Both facts have a simple
physical interpretation.  Let us consider, without loss of generality, the case
when the total energy of the process is of the order of the masses $M_i \sim
M$,
see previous sections.  Then the typical time, $\tau_p$, of the duration of the
production stage, as well as of the decay stages, is of order $1/M$.  This time
$\tau_p$ is much less than the characteristic time $\Delta t$ between the
various stages
\begin{equation}
\Delta t \sim {\rm max} \left( \frac{1}{\Gamma_i}, \frac{1}{\Gamma_j} \right)
\sim \frac{1}{\Gamma} .
\end{equation}
As a consequence the relative phases of photon/gluon emissions (or between
emission and absorption) at the different stages of the process are
approximately equal to $|k^0|\Delta t \sim |k^0|/\Gamma$.  When $|k^0| \gg
\Gamma$ this phase shift is large and therefore the radiative interference
effects are suppressed.

The second basic fact, the absence of infrared divergences in totally inclusive
cross-sections, has also a clear physical interpretation.  Infrared divergences
appear when we use states containing a definite number of photons/gluons.  Now
the acceleration of charge/colour leads to radiation with finite spectral
intensity at zero frequency and so the scattering or the creation of
charged/coloured particles is accompanied by the emission of an infinite number
of photons/gluons.  To obtain a physically meaningful cross section we must
include arbitrary numbers of emitted photons or gluons (in the case of gluons
we
need also to average over the initial colour states).

For the proof we only need the absence of those infrared divergences in the
total cross-section which are connected with the radiative interference between
the various production and decay stages.  To show this we first note that the
total cross-section is proportional to the imaginary part of the forward
scattering amplitude.  When this amplitude is expressed as the sum of Feynman
diagrams, all particles except the initial particles, appear as internal lines.
Therefore the interference photon/gluon lines must be attached to internal
lines, at least at one end.  But it is well known that infrared divergent
contributions only arise from photons/gluons with lines which couple at both
ends to external lines corresponding to on-mass-shell particles.  Because of
the
absence of infrared divergences the contribution from the region of small
photon/gluon energies $|k^0| \lapproxeq \Gamma$ is small due to the lack of
phase space.  On the other hand, as shown above, for $|k^0| \gg \Gamma$ the
interference effects are small due to the large time separations between the
various stages.  Therefore radiative interference between the production and
decay stages is suppressed by at least a factor $\Gamma/M$.  For multiple
exchanges the conclusion remains valid.  In the case of multiple exchange large
energies $k^0_i$ of individual photons/gluons are allowed, since the constraint
that the invariant mass of the unstable particle must not be shifted far from
its resonant value only requires
\begin{equation}
\mid\sum_i k^0_i\mid \lapproxeq \Gamma
\end{equation}
where the sum is performed over photons or gluons emitted $(k^0_i > 0)$ and
absorbed $k^0_i < 0)$ at one of the decay stages.  But because of the absence
of
infrared divergences there is again a suppression of at least one factor of
$\Gamma/M$ due to the restriction of the phase space imposed by (30).

Finally let us consider the Coulomb radiative effects.  Now there are infrared
singularities connected with the Coulomb interaction of charged/coloured
particles which are special in the sense that they are not cancelled by real
emissions, but rather they appear in matrix elements as a phase factor with an
infinite phase.  Therefore they do not appear in the expression for the cross
section and are not seen at all in the approach where we express the cross
section in terms of the forward scattering amplitude.

However there is a Coulomb interaction which is connected with small
photon/gluon frequencies, but which is not infrared divergent.  For two
charged/coloured particles, with reduced mass $\mu$ and momentum ${\bf q}$ in
their c.m.\ frame, the essential energies $k^0_c$ of the exchanged Coulombic
photons/gluons are typically $|k^0_c| \sim {\bf q}^2/\sqrt{({\bf q}^2 +
\mu^2)}$.  When ${\bf q}^2 \lapproxeq \mu\Gamma$ these energies are $|k^0_c|
\lapproxeq \Gamma$.  Thus for two slowly moving charged/coloured particles in
their c.m.\ frame there is an important Coulombic radiative interaction coming
from the region of small photon/gluon energies.  At first sight this appears to
violate our previous statement, that the contribution from the region $|k^0|
\lapproxeq \Gamma$ is small.  But that statement referred to interference
photons/gluons and it remains correct for them.  The reason is that for
interference between the different production and decay stages, the only
important Coulomb interactions are those between a charged/coloured decay
product of one of the unstable particles and some other particle (e.g.\ another
unstable particle or one of its decay products) with the interacting pair
slowly
moving in their c.m.\ frame (that is ${\bf q}^2 \lapproxeq \mu\Gamma$).  This
corresponds to a very small region of the available phase space and so these
Coulomb effects are also suppressed by at least a factor $\Gamma/M$.  This
concludes the proof of the theorem.

Some words should be added to explain the modifications $\delta(R,C)
\rightarrow
\delta(R,\bar{C})$ in going from  formula (26) for \lq\lq stable" heavy
particles to the realistic formula (27) for unstable particles.  Away from the
heavy particle production threshold the typical heavy particle interaction time
is $1/\sqrt{s} \lapproxeq 1/M$, i.e.\ much smaller than their lifetimes.  Thus
the influence of instability on the Coulomb corrections at the production stage
gives effects of relative order $\Gamma/M$ or less.  Thus $\delta(R,\bar{C})
\approx \delta(R,C)$.

The situation is different for heavy unstable particle production near
threshold.  Then the typical Coulomb interaction time $\tau_c$ can be
comparable
to, or even larger than, the particle's lifetime $\tau$
\begin{equation}
\tau_c \sim \frac{1}{k^0_c} \sim \frac{\mu}{{\bf q}^2} \gapproxeq
\frac{1}{\Gamma} = \tau .
\end{equation}
Therefore the Coulomb part $C$ of the radiative correction $\delta$ shown in
(26) will be considerably modified by instability, and hence it is denoted by
$\bar{C}$ in (27).  The calculations$^{[39,55]}$ of the modified contribution
$\bar{C}$ close to the threshold can be best done using old-fashioned
non-relativistic perturbation theory.  The diagrams are the same as in the
stable particle case, but we require the momentum $p_i = (\epsilon_i,{\bf
p}_i)$
of $A_i$ to satisfy $p^2_i = s_i$, and we must replace the energies of the
unstable particles by
\begin{equation}
\epsilon \longrightarrow \epsilon + \frac{iM\Gamma}{2\sqrt{{\bf p}^2 + M^2}}
\end{equation}
in the energy denominators of all intermediate states.

Concluding this section, let us make some comments concerning the effect of the
$W$ width on the Coulomb corrections to the process $e^+e^- \rightarrow W^+W^-
\, ^{[55,56]}$.  Accurate knowledge of these corrections is of importance for
the precise energy scan of $W^+W^-$ production through the threshold region.
In
principle, this provides us with an  \lq\lq interconnection-free" method of
measurement of the $W$ boson mass and width.  To my knowledge, modification of
the Coulomb interaction induced by the $W$ width effects was first studied in
Ref.\ [55].  In this paper the non-relativistic technique was used with the
virtuality of the $W$'s properly taken into account.

Consider the Coulomb attraction between the slowly moving $W$ bosons.  Because
the underlying Coulomb physics is different from the other radiative
corrections
it is possible to treat the Coulomb corrections separately from the
short-distance effects (see [58] and later Refs.\ [40,41]).  It was originally
discovered in QED$^{[55]}$ that when oppositely charged particles have low
relative velocity, $v \ll c$, Coulomb effects enhance the cross section by a
factor which, to leading order in $\alpha/v$, is $(1 + \alpha\pi/v)$, provided
the particles are stable.  Now it has been shown$^{[39,40]}$ that the Coulomb
effects may be radically modified when the interacting particles are
short-lived
rather than stable.  This is the case for $W$ bosons.  We would anticipate the
modification to be significant when the characteristic distance of the Coulomb
interaction $(d_C \sim \frac{1}{p})$ is greater than the typical spatial
separation when the diverging $W$ bosons decay $(d_{\tau} \sim
\frac{p}{m_W\Gamma_W})$.  The more that $d_{\tau}$ is less than $d_C$ the more
we expect the Coulomb attraction to be suppressed.  If we note that $p \approx
\sqrt{Em_W}$ then we see that the condition $d_{\tau} \lapproxeq d_C$
translates
into $E \lapproxeq \Gamma_W$ where $E = \sqrt{s} - 2m_W$ is the
non-relativistic
energy of the $W$ bosons.

Recall here that the interplay between the Breit-Wigner propagators and the
phase space factor (see Section 2) leads to larger values of $\langle p\rangle$
and, thus, induces an additional suppression of the Coulomb effects.

Allowing for the virtuality and the  finite width of the $W$ bosons we find the
following expression for the Coulomb correction in the non-relativistic
approximation$^{[55]}$:
\begin{equation}
\frac{\alpha}{v} \delta_{Coul} \simeq 2 {\rm Re} \left\{ \int
\frac{d^3k}{(2\pi)^3k^2} \cdot \frac{4\pi\alpha}{[({\bf p+k})^2/M_W - E -
i\Gamma_W]} \right\} ,
\end{equation}
where $p$ is the momentum of a virtual $W$, and $v = \frac{4p}{\sqrt{s}}$ is
the
relative velocity of the virtual $W$ bosons,
\begin{equation}
v = 2((\hat{s} - s_1 - s_2)^2 - 4s_1s_2)^{\frac{1}{2}}/s .
\end{equation}
On integrating (34) over $d^3k$ we find
\begin{equation}
\delta_{Coul} = 2 {\rm Re} \left\{ -i \, ln \left( \frac{p_1 - ip_2 + ip}{p_1 -
ip_2 - ip} \right) \right\} ,
\end{equation}
where $p_1 - ip_2 = \sqrt{M_W(-E - i\Gamma_W)}$, that is
\begin{equation}
p_{1,2} = \left[ M_W \left( \sqrt{E^2 + \Gamma^2_W} \mp E \right)/2
\right]^{\frac{1}{2}} .
\end{equation}
The result (35) giving the Coulomb correction for unstable particles should be
contrasted with the leading order value $(\delta^{(1)}_{Coul})_{st} = \pi$ for
stable particles.  To calculate the Coulomb corrections for interactions
between
stable particles it becomes increasingly necessary to include higher order
terms
in $\alpha/v$ as the threshold is approached, that is as $v \rightarrow 0$.
Summation of all the order $\left( \frac{\alpha}{v} \right)^n$ terms gives the
famous Sommerfeld-Sakharov result$^{[53]}$
\begin{displaymath}
(\delta_{Coul})_{st} = \frac{2\pi}{1 - e^{-Z}} ,
\end{displaymath}

\begin{equation}
Z = \frac{2\alpha\pi}{v} .
\end{equation}
It is worth noting that at the very threshold $(\delta_{Coul})_{st} =
2(\delta^{(1)}_{Coul})_{st}$.\footnote{This old result seems to have been
completely forgotten in many recent publications.}  However, as was first
indicated in [55] in the finite width case the instability of the particles
prevents the large distance effects from contributing and so the effects of the
higher-order Coulomb corrections are small.  One can roughly estimate that in
the presence of the width the expansion parameter in the Coulomb series is
\begin{displaymath}
\frac{\Delta\sigma_{Coul}}{\sigma} \sim \frac{\alpha\pi}{\sqrt{(E^2 +
\Gamma^2_W)^{\frac{1}{2}}/M_W}} < 0.15
\end{displaymath}
instead of $Z$ in the stable case.

Note that the contribution of the higher-order Coulomb effects can be exactly
calculated (if necessary) using the non-relativistic Green's function formalism
of Refs.\ [20,40,59].

Finally I would like to mention a claim of Ref.\ [56] that the non-relativistic
approach is not sufficient for the precise description of the Coulomb effects
in
the whole LEP2 region.  My personal feeling is that we need some further
studies
of this important issue.

\vspace*{0.6cm}

\noindent  {\bf 4.  Summary and Outlook}

\vspace*{0.4cm}

The large width $(\Gamma \sim O$(1 GeV)) of heavy unstable particles controls
the radiative interferences between emission occuring at different stages of
the
production processes.  The QCD interferences may efface the separate identities
of these particles and produce hadrons that cannot be uniquely assigned to
either of them.  Here we concentrated mainly on two topical problems, namely
the
QCD interconnection phenomena in events of the type $e^+e^- \rightarrow W^+W^-
\rightarrow q_1\bar{q}_2q_3\bar{q}_4$ and $e^+e^- \rightarrow t\bar{t}
\rightarrow bW^+\bar{b}W^-$.  We have shown$^{[23,24]}$ that, on the
perturbative level, these interference effects are suppressed and we have
applied hadronization models to help us estimate the nonperturbative effects.

One of the favoured methods of the $W$ and top mass determination is to
reconstruct them event by event.  The total contribution to the systematic
error
on the $W$ mass reconstruction was estimated conservatively as 40 MeV and the
uncertainty in the top mass was found on the level of less than 100 MeV.  We
believe that with sophisticated analysis methods these uncertainties can be
reduced.   Therefore, one may expect that in the foreseeable future the precise
determinations of $m_W$ and $m_t$ are not jeopardized by the QCD
interconnection
effects.

In some sense, the interconnection effects discussed here could be considered
as
only the tip of the iceberg.  Colour reconnection can  occur in any process
which involves the simultaneous presence of more than one colour singlet.  Many
of the techniques developed in Refs.\ [20,23,24] could be directly applied to
these problems.

Among other examples of practical importance are $e^+e^- \rightarrow Z^0H^0$,
$e^+e^- \rightarrow Z^0Z^0$, $pp/\bar{p}p \rightarrow W^+W^-$, $pp/\bar{p}p
\rightarrow t\bar{t}$, $pp/\bar{p}p \rightarrow t\bar{b}$, $pp/p\bar{p}
\rightarrow W^{\pm}H^0$, etc.  One could discuss also interferences with beam
jets.  The problem with these processes is that there are too many other
uncertainties which make systematic studies look very difficult.

QCD interconnection is interesting in its own right, since it potentially
provides a laboratory for a better understanding of the hadronization dynamics.
With a lot of hard work (and good luck), LEP2 could probably be in a position
to
discriminate between some hadronization models.

The other methods of the $W$ mass measurements$^{[10,11]}$ also require a clear
understanding of the role of the $W$ width, this time because of the QED
radiative phenomena.  Thus, the width effects modify the QED Coulomb
corrections
to the cross-section of the process $e^+e^- \rightarrow W^+W^-$, which should
be
known with a high accuracy for the measurements scanning across the $WW$
threshold region$^{[55,56]}$.  The $W$ width can have an impact on the
measurements of $m_W$ from the shape of the lepton spectrum (the lepton
end-point method).  To my knowledge this study has never been systematically
addressed.

At last, a direct determination of the $W$ width itself using the transverse
mass distribution of $W \rightarrow e\nu$ decays$^{[9]}$ requires, in
principle,
a careful analysis of the QED interconnection effects.

We cannot today predict what will come out of the forthcoming systematic
studies, e.g., within the working groups for LEP2 and a future $e^+e^-$ linear
collider.  The results reviewed here can be considered as a starting point for
more refined and detailed investigations.  We believe that radiative
interference phenomena will be of topical interest for many years to come.

\vspace*{0.6cm}

\noindent {\bf Acknowledgements}

\vspace*{0.4cm}

It is a pleasure for me to thank Risto Orava and Masud Chaichian for creating
such a stimulating atmosphere and for the warm hospitality in Lapland.  I wish
to thank V.S.\ Fadin, C.J.\ Maxwell, W.J.\ Stirling, N.G.\ Uraltsev and
especially T.\ Sj\"{o}strand for fruitful discussions.  This work was supported
by the United Kingdom Particle Physics and Astronomy Research Council.

\vspace*{0.6cm}

\noindent {\bf  References}

\vspace*{0.4cm}
\begin{enumerate}
\item  Yu.\ Galaktionov, these proceedings.
\item  A.\ Blondel, Lectures at SLAC Summer Institute, \lq\lq Particle Physics,
Astrophysics and Cosmology", Aug.\ 1994.
\item  R.\ Dubois, ibid.
\item  CDF Collaboration, F.\ Abe et al., {\it Phys.\ Rev.\ Lett} {\bf 73}
(1994) 225; {\it Phys.\ Rev.} {\bf D50} (1994) 2966.
\item  L.B.\ Okun, these proceedings.
\item  For a recent review see, e.g., B.A.\ Kniehl, KEK preprint, KEK-TH-412,
1994.
\item  L.\ Nodulman, Argonne preprint, ANL-HEP-CP-94-51.
\item  M.\ Abolins, Lecture at SLAC Summer Institute, \lq\lq Particle Physics,
Astrophysics and Cosmology", Aug.\ 1994.
\item  CDF Collaboration, F.\ Abe et al., FERMILAB-PUB-94/301-E (1994).
\item  LEP2 workshop presentation by L.\ Camilleri at the LEPC open meeting,
CERN, November 1992; D.\ Treille, CERN preprint, CERN-PPE/93-54/REV; J-E.\
Augustin, CERN preprint, CERN-PPE/94-81.
\item S.\ Katsanevas, talk at the Schwarzwald Workshop 94.
\item  R.\ Orava, A.\ Skrinsky, V.\ Telnov, these proceedings.
\item  $e^+e^-$ Collisions at 500 GeV: The Physics Potential, ed.\ P.M.\
Zerwas,
DESY 92-123, Parts A,B and C.
\item  Physics and Experiments with $e^+e^-$ Linear Colliders, eds.\ R.\ Orava,
P.\ Eerola and M.\ Nordberg (World Scientific, Singapore, 1992).
\item  Physics and Experiments with Linear $e^+e^-$ Colliders, eds.\ F.A.\
Harris, S.L.\ Olsen, S.\ Pakvasa and X.\ Tata (World Scientific, 1993).
\item V.A.\ Khoze, L.H.\ Orr and W.J.\ Stirling, {\it Nucl.\ Phys.} {\bf B378}
(1992) 413.
\item  Yu.L.\ Dokshitzer et al., {\it Nucl.\ Phys.} {\bf B403} (1993) 65.
\item  V.A.\ Khoze, J.\ Ohnemus and W.J.\ Stirling, {\it Phys.\ Rev.} {\bf D78}
(1994) 1237.
\item  V.S.\ Fadin, V.A.\ Khoze and A.D.\ Martin, {\it Phys.\ Lett.} {\bf B320}
(1994) 141.
\item  V.S.\ Fadin, V.A.\ Khoze and A.D.\ Martin, {\it Phys.\ Rev.} {\bf D79}
(1994) 2247.
\item  Y.\ Sumino, Tohoku University preprint, TU-469 (1994); K.\ Fujii, T.\
Matsui and Y.\ Sumino, {\it Phys.\ Rev.} {\bf D50} (1994) 4341.
\item  J.H.\ K\"{u}hn in Ref.\ [15], vol.\ 1, p.\ 72.
\item  T.\ Sj\"{o}strand and V.A.\ Khoze, {\it Z.\ Phys.} {\bf C62} (1994) 281;
{\it Phys.\ Rev.\ Lett.} {\bf 72} (1994) 28.
\item  V.A.\ Khoze and T.\ Sj\"{o}strand, {\it Phys.\ Lett.} {\bf B328} (1994)
466.
\item  G.\ Gustafson, U.\ Pettersson and P.\ Zerwas, {\it Phys.\ Lett.} {\bf
B209} (1988) 90.
\item  G.\ G\"{u}stafson and J.\ H\"{a}kkinen, Lund preprint, LU TP 94-9.
\item  Yu.L.\ Dokshitzer, V.A.\ Khoze, A.H.\ Mueller and S.I.\ Troyan, \lq
Basics of Perturbative QCD', ed.\ J.\ Tran Thanh Van (Editions Frontieres,
Gif-sur-Yvette, 1991).
\item  Yu.L.\ Dokshitzer et al., {\it Rev.\ Mod.\ Phys.} {\bf 60} (1988) 373.
\item  Ya.I.\ Azimov et al., {\it Phys.\ Lett.} {\bf 165B} (1985) 147; Yu.L.\
Dokshitzer, V.A.\ Khoze and S.I.\ Troyan, {\it Sov.\ J.\ Nucl.\ Phys.} {\bf 50}
(1990) 505.
\item  See first Ref.\ in [23].
\item  B.\ Andersson et al., {\it Phys.\ Rep.} {\bf 97} (1983) 31.
\item  T.\ Sj\"{o}strand and M.\ Bengtsson, {\it Comp.\ Phys.\ Commun.} {\bf
43}
(1987) 367; H.-U.\ Bengtsson and T.\ Sj\"{o}strand, {\it Comp.\ Phys.\ Commun.}
{\bf 46} (1987) 43; T.\ Sj\"{o}strand, preprint CERN-TH.6488/92.
\item  Y.\ Sumino et al., {\it Phys.\ Rev.} {\bf D47} (1992) 56; M.\
Je\.{z}abek, J.H.\ K\"{u}hn and T.\ Teubner, {\it Z.\ Phys.} {\bf C56} (1992)
653; M.\ Je\.{z}abek and T.\ Teubner, {\it Z.\ Phys.} {\bf C59} (1993) 669.
\item  J.D.\ Bjorken, S.J.\ Brodsky and H.J.\ Lu, {\it Phys.\ Lett.} {\bf B286}
(1992) 153; H.J.\ Lu, S.J.\ Brodsky and V.A.\ Khoze, {\it Phys.\ Lett.} {\bf
B312} (1993) 215.
\item  P.\ Igo-Kemenes et al., in Ref.\ [13], Part C, p.\ 319; K.\ Fujii, KEK
preprint 94-38; Y.\ Sumino in Ref.\ [15], v.\ II, p.\ 439.
\item  For a recent review see, e.g., M.\ Je\.{z}abek, Karlsruhe preprint TTP
94-09.
\item  J.H.\ K\"{u}hn, {\it Acta Phys.\ Aust.\ Suppl.} {\bf 24} (1982) 203.
\item  I.I.\ Bigi et al., {\it Phys.\ Lett.} {\bf B181} (1986) 157.
\item  V.S.\ Fadin and V.A.\ Khoze, {\it JETP Lett.} {\bf 46} (1987) 525; {\it
Sov.\ J.\ Nucl.\ Phys.} {\bf 48} (1988) 309; {\bf 53} (1991) 692.
\item  V.S.\ Fadin and V.A.\ Khoze, Proc.\ of 24th LNPI Winter School,
Leningrad, Vol.\ 1, p.\ 3 (1989).
\item  V.S.\ Fadin, V.A.\ Khoze and T.\ Sj\"{o}strand, {\it Z.\ Phys.} {\bf
C48}
(1990) 613.
\item  E.C.\ Poggio, H.R.\ Quinn and S.\ Weinberg, {\it Phys.\ Rev.} {\bf D13}
(1976) 1958.
\item  T.\ Sj\"{o}strand and P.\ Zerwas, in [13], Part A, p.\ 463.
\item  I.I.\ Bigi et al., {\it Phys.\ Rev.} {\bf D50} (1994) 2234.
\item  M.\ Beneke and V.M.\ Braun, {\it Nucl.\ Phys.} {\bf 426} (1994) 301; M.\
Beneke, V.M.\ Braun and V.I.\ Zakharov, MPI report, MPI-PhT/94-19.
\item  B.H.\ Smith and M.B.\ Voloshin, Univ.\ Minnesota reports,
UMN-TH-1241/94;
UMN-TH-1252/94.
\item  S.\ Fanchiotti, B.\ Kniehl and A.\ Sirlin, {\it Phys.\ Rev.} {\bf D48}
(1993) 307.
\item  N.\ Gray et al., {\it Z.\ Phys.} {\bf C48} (1990) 673.
\item  For a review see, A.H.\ Mueller in \lq\lq QCD - Twenty Years Later",
Proc.\ Int.\ Conf., Aachen, 1992, eds.\ P.\ Zerwas and H.\ Kastrup (World
Scientific, Singapore, 1993) Vol.\ 1, p.\ 162.
\item  M.\ Je\.{z}abek, J.H.\ K\"{u}hn and T.\ Teubner in [13], Part C, p.\
303.
\item  A.\ Sirlin, New York Univ.\ preprint NYU-TH-94/11/02.
\item  For a review see, e.g., J.L.\ Lopez, CTP-TAMU-39/94.
\item  A.\ Sommerfeld, \lq\lq Atombau und Spektrallinien", Bd.\ 2 (Vieweg,
Braunschweig, 1939); A.D.\ Sakharov, JETP {\bf 18} (1948) 631.
\item M.J.\ Strassler and M.E.\ Peskin, {\it Phys.\ Rev.} {\bf D43} (1991)
1500.
\item  V.S.\ Fadin, V.A.\ Khoze and A.D.\ Martin, {\it Phys.\ Lett.} {\bf B311}
(1993) 31.
\item  D.\ Bardin, W.\ Beenakker and A.\ Denner, {\it Phys.\ Lett.} {\bf B317}
(1993) 213.
\item  K.\ Melnikov and O.\ Yakovlev, {\it Phys.\ Lett.} {\bf B324} (1994) 217.
\item I.\ Harris and L.M.\ Brown, {\it Phys.\ Rev.} {\bf 105} (1957) 1656.
\item  V.S.\ Fadin, V.A.\ Khoze and M.I.\ Kotsky, {\it Z.\ Phys.} {\bf C64}
(1994) 45.
\end{enumerate}

\end{document}